%% file: main.tex
\documentclass[acmtog]{acmart}
\acmSubmissionID{1485}

\usepackage{booktabs} % For formal tables

\usepackage[table]{xcolor}

\definecolor{tabfirst}{rgb}{1.0, 0.7, 0.7}
\definecolor{tabsecond}{rgb}{1.0, 0.85, 0.7}
\definecolor{tabthird}{rgb}{1.0, 1.0, 0.7}
\newcommand{\first}[1]{\cellcolor{tabfirst}\textbf{#1}}
\newcommand{\second}[1]{\cellcolor{tabsecond}#1}
\newcommand{\third}[1]{\cellcolor{tabthird}#1}

\usepackage[ruled]{algorithm2e} % For algorithms

\SetAlFnt{\small}
\SetAlCapFnt{\small}
\SetAlCapNameFnt{\small}
\SetAlCapHSkip{0pt}

\acmJournal{TOG}
\setcopyright{cc}
\setcctype{by}
\acmJournal{TOG}
\acmYear{2026} \acmVolume{45} \acmNumber{6} \acmArticle{200}
\acmMonth{12} \acmDOI{10.1145/3842527}

\begin{document}
% Title portion
\title{Inverse Rendering for Modeling with Line Primitives}

% DO NOT ENTER AUTHOR INFORMATION FOR ANONYMOUS TECHNICAL PAPER SUBMISSIONS TO SIGGRAPH 2019!
\author{Kenji Tojo}
\orcid{0000-0001-9415-0701}
\affiliation{%
 \institution{ETH Zürich}
 \city{Zürich}
 \country{Switzerland}}
\email{ketojo@ethz.ch}

\author{Ariel Shamir}
\orcid{0000-0001-7082-7845}
\affiliation{%
 \institution{Reichman University}
 \city{Herzliya}
 \country{Israel}}
\email{arik@runi.ac.il}

\author{Nobuyuki Umetani}
\orcid{0000-0003-1251-970X}
\affiliation{%
 \institution{The University of Tokyo}
 \city{Tokyo}
 \country{Japan}}
\email{n.umetani@gmail.com}

\author{Bernd Bickel}
\orcid{0000-0001-6511-9385}
\affiliation{%
 \institution{ETH Zürich}
 \city{Zürich}
 \country{Switzerland}}
\email{bickelb@ethz.ch}

\renewcommand\shortauthors{Tojo et al.}

\begin{teaserfigure}
\includegraphics{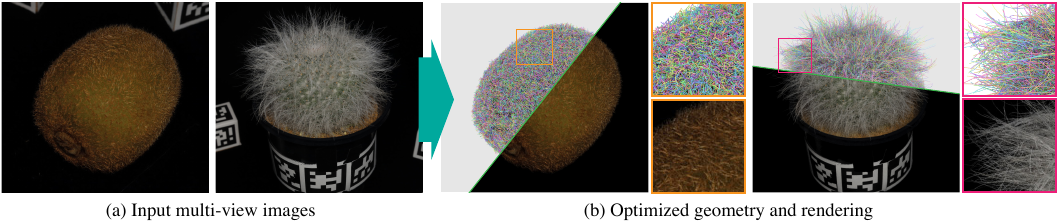}
\caption{%
(a) Given a set of input images, (b) our method faithfully reconstructs highly detailed fuzzy geometry composed of many thin fibers as a collection of explicit line primitives.
Both training and rendering remain directly compatible with efficient rasterization pipelines.
}
\Description[teaser figure]{teaser figure}
\label{fig:teaser}
\end{teaserfigure}

\input{sec/0_abstract}





\maketitle

\input{sec/1_intro}
\input{sec/2_related}

\input{sec/3_preliminary}

\input{sec/4_method}

\input{sec/5_results}

\input{sec/6_conclusion}

% DO NOT INCLUDE ACKNOWLEDGMENTS IN AN ANONYMOUS SUBMISSION TO SIGGRAPH 2019
\begin{acks}

We thank the anonymous reviewers for their valuable feedback.
This work was partly conducted while the first author was at the University of Tokyo, where he was supported by JSPS KAKENHI Grant Number 23KJ0699.

\end{acks}

% Bibliography
\bibliographystyle{ACM-Reference-Format}
\bibliography{main}

\input{sec/X_suppl}

\end{document}

%% file: sec/0_abstract.tex
\begin{abstract}
Faithfully capturing diverse real-world objects with fuzzy, anisotropic structures—such as hair, fur, fibers, and textiles—for efficient real-time visualization remains challenging.
Recent radiance field reconstruction methods capture these structures from multi-view images using translucent volumetric primitives such as 3D Gaussians rather than opaque low-dimensional primitives (e.g., triangles, line segments, and polylines), thereby limiting compatibility with standard depth-tested rasterization, reflection modeling, and physical simulation.
We present an inverse rendering method for reconstructing fuzzy geometry using explicit line segments, which are rasterized on a subpixel grid for anti-aliasing to reproduce a semi-transparent appearance.
While straightforward to render, optimizing numerous line primitives to match target images poses a significant challenge.
We address this by introducing a stochastic differentiable rasterizer for line segments that produces informative gradients with respect to vertex positions, attributes, and discrete connectivity.
Experiments on synthetic and real-world datasets show that our method outperforms surface-based approaches in capturing fuzzy boundaries and achieves quality comparable to volumetric representations while relying entirely on explicit geometry.
The resulting representation integrates seamlessly with standard graphics pipelines, enabling cross-platform rendering, various shading models, and physical simulation.
\end{abstract}

%% file: sec/1_intro.tex
\section{Introduction}
\label{sec:intro}

Faithfully representing the rich visual appearance of real-world objects remains a central challenge in computer graphics.
Beyond visual fidelity, many applications—such as games, virtual and augmented reality, telepresence, and robotics—require geometry representations that can be efficiently rendered, analyzed, edited, and simulated within standard graphics pipelines across diverse hardware platforms.
Consequently, most existing approaches rely on surface-based representations, modeling objects as surface meshes augmented with texture maps encoding fine-scale appearance and geometric detail (e.g., albedo, normals, and displacement).

However, such explicit representations fundamentally assume that object geometry can be represented as locally smooth surfaces with parameterizable detail.
This assumption breaks down for fuzzy or weakly defined structures, such as hair, fur, fibers, and other anisotropic materials, where boundaries are inherently ambiguous, and appearance emerges from the aggregation of many thin elements.
Capturing these phenomena with surfaces alone is difficult and often requires expensive procedural detail generation.
As a result, a broad class of real-world objects—including plants, animals, food, and textiles—remains challenging to represent faithfully within traditional surface-based frameworks.

Recent advances in view synthesis address this limitation by shifting the representation paradigm from explicit surfaces to implicit volume representation.
Methods such as NeRF~\cite{mildenhall2020nerf} model scenes as continuous emissive fields, naturally capturing fuzzy boundaries and complex appearance.
Building on this, 3D Gaussian splatting~\cite{kerbl3Dgaussians} represents scenes using volumetric primitives, enabling real-time rendering and intuitive editing.
While these approaches excel at representing ambiguous and fine-scale structures, they do so by abandoning explicit low-dimensional geometric structure.
Such implicit representation is difficult to integrate into standard graphics pipelines that rely on depth-tested rasterization, geometric reasoning, and physical simulation.

This tension raises a fundamental question: how can we represent the appearance of fuzzy, weakly defined, and anisotropic geometry using structured explicit primitives without sacrificing fidelity?
In this work, we propose an inverse rendering method that bridges this gap by modeling diverse fuzzy, fiber-like structures as a collection of line primitives (e.g., line segments and polylines), an explicit 1D representation that captures fine-scale structure while remaining compatible with traditional rasterization.
Specifically, objects are described as 3D vertices connected into polylines, which are rasterized on a subpixel grid and anti-aliased using multi-sample anti-aliasing (MSAA).
This produces a semi-transparent appearance through the aggregation of many opaque line segments, enabling the faithful reproduction of fuzzy boundaries and anisotropic effects entirely within a standard graphics pipeline (Figure~\ref{fig:teaser}).

This line primitive representation offers several key advantages.
First, it provides a \emph{compact} description of thin geometric elements, making it naturally suited for representing fibers, strands, hair, and other elongated structures.
Second, its \emph{explicit} representation remains compatible with established rendering techniques, including z-buffer-based depth testing, shading models, and hardware acceleration.
Finally, it preserves the \emph{structure} of the underlying thin geometry—such as tangent directions and polyline connectivity—which can be leveraged for shading and physical simulation.

While the rendering process itself is straightforward, optimizing a large set of line primitives to reproduce complex target appearances remains highly challenging.
To address this, we present a fully differentiable rasterization framework for line primitives that produces informative gradients with respect to vertex positions, attributes, and even discrete connectivity.
We build on stochastic differentiable rasterization techniques for triangle primitives~\cite{tojo2026diffsoup} and extend them to subpixel-width line segments with high-quality anti-aliasing, while supporting a range of shading models via vertex attributes, including direct radiance parameterizations using spherical harmonics.
This enables dynamic optimization in which line primitives can appear, move, connect, and disappear, faithfully reproducing complex fuzzy structures and fine details.

We evaluate our approach on both synthetic and real-world datasets by reconstructing line-based representations from multi-view images.
We compare our method with state-of-the-art baselines, including layered surfaces~\cite{Esposito2025VolSurfs}, neural radiance fields~\cite{adaptiveshells2023}, and 3D Gaussian splatting~\cite{kerbl3Dgaussians}.
Compared with surface-based methods, our representation more effectively captures fuzzy and anisotropic appearance.
At the same time, it achieves reconstruction quality comparable to volumetric approaches, while relying entirely on explicit, opaque geometry.
Beyond fidelity, we demonstrate seamless integration with standard graphics pipelines, including cross-platform deployment, advanced shading models, physics-based animation, and integration into surface-based scenes.
Our open-source implementation and real-world capture dataset are available through \textcolor{ACMDarkBlue}{\url{https://github.com/kenji-tojo/inverse-line-primitives}}.

%% file: sec/2_related.tex
\section{Related Work}
\label{sec:related}

\begin{figure*}[t]
\centering
\includegraphics{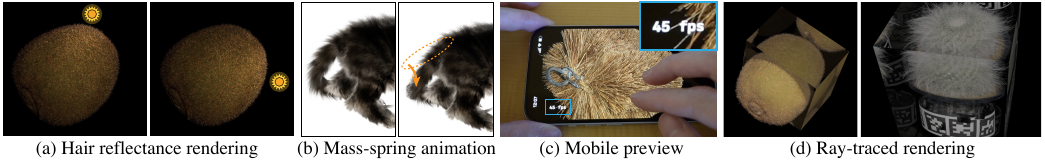}
\caption{%
Applications enabled by our reconstructed line-based representations.
(a) Strand reflectance rendering using a physically inspired hair scattering model~\cite{marschner2003}.
(b) Mass-spring animation with KNN-augmented connectivity between reconstructed polylines.
(c) Interactive mobile preview using up to degree-1 spherical harmonics.
(d) Exporting explicit line primitives as constant-radius cylinders for offline ray-traced rendering.
The supplemental video demonstrates applications (a), (b), and (c).
See Section~\ref{sec:results} for further discussion.
}
\label{fig:applications}
\Description[Applications]{Applications}
\end{figure*}

\paragraph*{Differentiable rendering}
Differentiable rendering~\cite{kato2020} solves inverse problems by computing gradients through image synthesis procedures, including light transport simulation~\cite{NimierDavidVicini2019Mitsuba2,Li:2018:DMC,Zhang:2019:DTRT}, rasterization~\cite{chen2019dibr,Loper:ECCV:2014,liu2019softras,Laine2020diffrast,Pidhorskyi2024rasterized}, vector graphics~\cite{Li:2020:DVG}, parametric geometry~\cite{Worchel2023parametric}, and programmable shaders \cite{Yang:2022:AAF,bangaru2023slangd}.
We build on the rasterization-based approach~\cite{Pidhorskyi2024rasterized} to efficiently reconstruct numerous line primitives using hardware rasterization, avoiding the computational cost of full light transport simulation.

Prior differentiable rendering methods for curves rely on analytic ray intersections~\cite{Zhang2023Projective}, sequential line segments~\cite{Tojo2024Wireart}, camera-facing quads~\cite{Takimoto_2024_CVPR}, or tubular meshes~\cite{Worchel2023parametric}, typically assuming a fixed topology and limiting adaptation to highly unstructured fuzzy geometry.
In contrast, we extend the recent stochastic triangle rasterization approach~\cite{tojo2026diffsoup} to anti-aliased subpixel-width line segments.
This enables line primitives to appear, disappear, and reorganize into explicit 1D manifolds from random initialization.

Differentiable rendering has also been applied to asset simplification, where fine structures such as leaves, hair, and fur are approximated using textured surface proxies~\cite{Hasselgren2021,Tojo2025Strands2Cards,zheng2025haircard,Esposito2025VolSurfs}.
However, representing dense microstructures with coarse surfaces with textures remains fundamentally difficult, as the proxy geometry no longer matches the underlying directional structures.
In contrast, we directly reconstruct such geometry using explicit line primitives, preserving fine structures while remaining efficient to rasterize.

\paragraph*{Radiance fields}
To capture the complex interplay between geometry and appearance in real-world scenes, neural radiance fields \cite{mildenhall2020nerf} (NeRFs) represent 3D scenes as continuous volumetric radiance fields.
A large body of work has extended this formulation to recover features associated with classical graphics pipelines, including anti-aliased rendering~\cite{barron2021mipnerf}, unbounded scenes with environment modeling~\cite{barron2022mipnerf360}, and efficient spatial representations via voxel grids and related structures~\cite{yu2022plenoxels,mueller2022instant,liu2020neural,Karnewar2022ReLUFields,adaptiveshells2023}.

More recently, 3D Gaussian Splatting~\cite{kerbl3Dgaussians} (3DGS) replaces the implicit radiance fields rendered by ray marching with 3D Gaussian primitives rendered by depth-order splatting.
While 3DGS enables real-time, high-fidelity rendering and intuitive scene manipulation, it remains fundamentally \emph{point-based} without explicit connectivity.
As a result, such representations are inherently incompatible with a large body of existing geometry processing and graphics algorithms defined on \emph{low-dimensional manifolds} (e.g., 2D surfaces or 1D curves), including depth-tested rendering, Laplacian-based editing~\cite{nealen2006laplacian}, geodesic distance computation~\cite{crane2017heat}, reflectance modeling~\cite{marschner2003}, physical simulation~\cite{bergou2008discrete}, and collision handling~\cite{li2021codimensional}.
This limitation extends to related splatting approaches based on 2D Gaussians~\cite{Huang2DGS2024}, convex primitives~\cite{Held20243DConvex}, and translucent triangles~\cite{Held2025Triangle}.

We address the fundamental tension between visual fidelity and explicit manifold structure for fuzzy and fiber-like geometry. 
Volumetric representations reproduce appearance but do not yield manifolds, while surface extraction methods recover geometry but struggle to efficiently capture their semi-transparent, fuzzy appearance~\cite{wang2021neus,Zhang2025Radiance,li2023neuralangelo,Huang2DGS2024,guedon2025milo,Esposito2025VolSurfs}.
We resolve this by reconstructing such regions using subpixel-width line segments, yielding an explicit 1D manifold whose fuzzy appearance emerges through screen-space filtering.
This enables faithful rendering while restoring compatibility with geometry-centric pipelines.

\paragraph*{Hair capture}
1D curve representations are widely used to model hair strands and fur, supported by dedicated reflectance models \cite{marschner2003} and physical simulation techniques~\cite{bergou2008discrete,li2021codimensional}.
Prior work on hair reconstruction typically estimates directional fields to guide strand generation~\cite{zhou2024groomcap,Jakob2009Capturing,paris08hair,shen2023ct2hair,Takimoto_2024_CVPR}.
Notably, Nam et al.~\shortcite{nam2019strand} construct accurate directional fields from a dense point cloud using strand-aware patch matching.
While directional fields are effective for inferring smoothly varying fiber orientation, these assumptions often break down for general fuzzy surfaces, where strands are highly unstructured and densely cluttered.
In contrast, we directly optimize explicit line primitives using differentiable rendering, without relying on directional fields.
As a result, our method supports more diverse objects than the differentiable rendering approach for hair reconstruction~\cite{Takimoto_2024_CVPR}, including animals, fabrics, and plants.

\paragraph*{1D representation}
1D curves and feature lines have long been used for visual representation and abstraction~\cite{benard2019line}, and more recently combined with differentiable rendering to generate line drawings and 3D abstractions~\cite{vinker2022clipasso,choi20243doodle,liu2025wir3d,Tojo2024Wireart}.
While these methods produce sparse, stylized representations that convey structure with few primitives, we optimize dense collections of line segments to faithfully reproduce rich, realistic 3D appearance.

\paragraph*{Mesh optimization}
Optimization of mesh topology has been studied in surface modeling~\cite{garland1997,hoppe1993mesh,liu2025simplifying,botsch2004}, fluid simulation~\cite{brochu2009}, and more recently in geometry learning~\cite{son2024dmesh,son2025dmesh++,spacemesh2024,shen2021dmtet,shen2023flexicubes}.
Several recent works optimize mesh connectivity during differentiable rendering~\cite{palfinger2022continuous,Held2025MeshSplatting,son2025dmesh++,son2024dmesh}.
While these methods operate on the 2D surface topology of triangle meshes, we instead optimize the 1D polyline topology, adaptively allocating line primitives based on opacity and geometry, within our differentiable rasterization framework.

%% file: sec/3_preliminary.tex
\section{Preliminary}
\label{sec:preliminary}
We aim to reconstruct fuzzy, high-detail appearance arising from many thin elements using explicit line primitives.
Unlike 3DGS \cite{kerbl3Dgaussians}, we rely on \emph{opaque} primitives, which are straightforward to integrate into standard rendering pipelines where visibility is resolved through z-buffering.

Although efficient to render, opaque primitives are significantly more challenging to optimize than volumetric approaches, since the discrete depth test blocks gradient propagation to occluded primitives.
To address this, we build on DiffSoup~\cite{tojo2026diffsoup}, a stochastic differentiable rasterizer that enables differentiation of binary opacity under discrete rasterization.
The remainder of this section reviews the DiffSoup rasterizer, which extends continuous \emph{stochastic surface} formulations~\cite{Zhang2025Radiance,zhang2025many} to discrete, object-order rasterization with depth testing.

\begin{figure*}[t]
\centering
\includegraphics{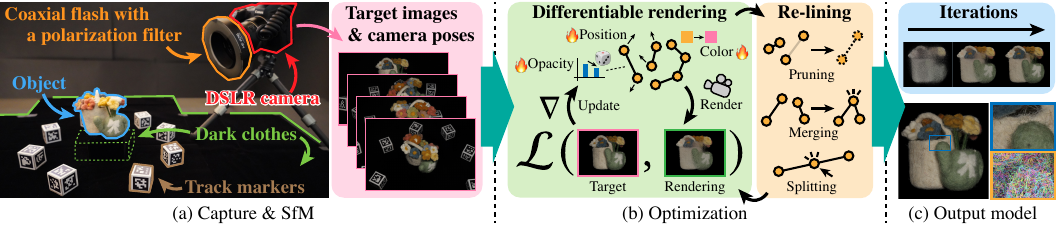}
\caption{%
Overview of our workflow.
(a) We capture multi-view images in a dark environment using a DSLR camera with a coaxial flash and polarization filter to suppress specular reflections and facilitate geometry reconstruction, then estimate camera poses using Structure-from-Motion~\cite{schonberger2016structure} (SfM).
(b) We optimize explicit line primitives through differentiable rendering with periodic discrete updates, termed \emph{re-lining}, including pruning, merging, and splitting.
(c) After optimization, our method reconstructs detailed line-based models that enable faithful multi-view rendering.
}
\label{fig:workflow}
\Description[Our workflow figure]{Our workflow}
\end{figure*}

\paragraph*{Depth testing for opaque primitives}
For each pixel, a rasterizer generates a set of fragments $\mathcal{F}$ by sampling on primitives and interpolating per-vertex attributes.
Each fragment $f\in\mathcal{F}$ carries a color $\mathbf{C}_f$ and a depth $d_f$.
For opaque primitives, the final pixel color $\hat{\mathbf{C}}= \mathbf{C}_{f^*}$ is given by the frontmost fragment $f^*$,
\begin{equation}
\label{eq:depth_test}
f^* = \underset{f\in\mathcal{F}}{\arg\min}\,d_f.
\end{equation}
The z-buffer method computes the frontmost fragment without explicit sorting, in contrast to the back-to-front alpha blending of fully sorted primitives (e.g., 3D Gaussians~\cite{kerbl3Dgaussians}).

\paragraph*{Stochastic rasterization of opaque primitives}
To differentiate this discrete fragment selection, DiffSoup assigns a continuous opacity value to each primitive.
This opacity is optimized as a continuous value but eventually converges to either 0 or 1, unlike~\cite{kv2025stochasticsplats}, which also uses stochastic rasterization but optimizes the \emph{translucent} appearance.
The visible fragment is stochastically selected based on primitive opacity as
\begin{equation}
\label{eq:stoc_depth_test}
f^* = \underset{f\in\mathcal{F}:\,\alpha_f>\tau_f}{\arg\min}\,d_f,
\end{equation}
where $\alpha_f\in[0,1]$ is the interpolated opacity at the fragment $f$, and $\tau_f\in[0,1]$ is a uniform random sample.
That is, each fragment is stochastically activated with probability $\alpha_f$ by testing whether $\alpha_f>\tau_f$.
The pixel color $\hat{\mathbf{C}}$ is now a random variable, where the probability $p_f$ of fragment $f$ being selected as $\hat{\mathbf{C}}=\mathbf{C}_f$ is
\begin{equation}
\label{eq:selection_prob}
p_f = \alpha_f \prod_{f'\in\mathcal{F}:\,d_{f'}<d_f}(1-\alpha_{f'}),
\end{equation}
mirroring the transmittance term used in alpha blending.

The $L_1$ color loss against the ground truth is defined as an average over stochastic renderings $\mathcal{L}=\mathbb{E}_{p_f}[\mathcal{L}_1(\mathbf{C}_f)]$.
By viewing the rasterizer as a \emph{sampler}, the gradient of $\mathcal{L}$ can be computed via the likelihood-ratio gradient estimator~\cite{williams1992simple,glynn1990likelihood} as
\begin{equation}
\label{eq:lr_gradient}
\nabla\mathcal{L} = \mathbb{E}_{p_f}[\nabla\mathcal{L}_1(\mathbf{C}_f)] + \mathbb{E}_{p_f}[\mathcal{L}_1(\mathbf{C}_f)\,\nabla\log p_f],
\end{equation}
where the first term is the color gradient computed via the standard differentiable rasterization~\cite{Pidhorskyi2024rasterized}, and the second is the score-function term that differentiates fragment existence.
The log-probability gradient can be evaluated without sorting
\begin{equation}
\label{eq:log_prob_grad}
\nabla_{\alpha_{f'}}\log p_f = \begin{cases}
1/\alpha_f & \text{if } f'=f, \\
-1/(1-\alpha_{f'}) & \text{if } d_{f'}<d_f,
\end{cases}
\end{equation}
where $\nabla_{\alpha_{f'}}$ denotes the gradient with respect to per-fragment opacity $\alpha_{f'}$, and the gradient is zero for all other fragments.
At test time, the threshold $\tau_f$ is set to $0.5$ for all fragments.
We refer interested readers to the original paper~\cite{tojo2026diffsoup} for further details.

%% file: sec/4_method.tex
\section{Method}
\label{sec:method}

While DiffSoup enables differentiable rendering with opaque primitives, it aims to represent scenes compactly using a small number of large triangles, relying on continuous textures to reproduce fine-scale detail.
As a result, it does not explicitly optimize for anti-aliased geometry, making it difficult to capture appearance dominated by subpixel geometric \emph{boundaries}, such as dense fur or fibers.
In this work, we treat visual detail as arising from many thin boundaries, represented by line primitives and rendered with anti-aliasing.
This enables faithful reconstruction of fuzzy, anisotropic appearance while remaining fully compatible with depth-tested graphics pipelines.
Our overall workflow is summarized in Figure~\ref{fig:workflow}.

\subsection{Scene Representation with Line Primitives}
Given target images $\mathcal{T}=\{\mathbf{T}_t\}_{t=1}^{\vert\mathcal{T}\vert}$ with corresponding camera poses, our goal is to reconstruct a 3D scene using explicit primitives defined by interpolated vertices.
Such representations are efficiently rendered via depth testing and are compatible with a range of geometry-processing and simulation algorithms.
A common choice is a triangle-based representation, which models appearance through barycentric interpolation over surface elements.
While effective for smooth surfaces, this formulation struggles when appearance is dominated by complex boundaries and anisotropic structures.
We instead represent such structures using line primitives, enabling explicit representation of complex boundary geometry.

Specifically, we represent the scene using vertex positions $\mathcal{V}=\{\mathbf{v}_v\}_{v=1}^{\vert\mathcal{V}\vert}$ and edge indices $\mathcal{E}=\{\mathbf{e}_e\}_{e=1}^{\vert\mathcal{E}\vert}$.
Each vertex is a 3D point $\mathbf{v}_v\in\mathbb{R}^3$, and each edge is a pair of vertex indices $\mathbf{e}_e=(v_1^e,v_2^e)\in\{1,\dots,\vert\mathcal{V}\vert\}^2$.
Although this representation accommodates a range of topologies, from disjoint line segments to wireframe meshes, we focus on \emph{polyline} structures without branches or junctions, where all vertices have valence 1 or 2.
Such polylines are efficiently rasterized on modern graphics hardware and support existing curve-processing techniques.

\paragraph*{Auxiliary parameters}
To model view-dependent appearance, we associate auxiliary attributes $\mathcal{A}=\{\mathbf{A}_v\}_{v=1}^{\vert\mathcal{V}\vert}$ with each vertex.
These may include learnable features, geometric quantities (e.g., tangents), or combinations thereof.
We further learn per-edge opacity values to enable emergence and extinction of primitives during optimization, which improves robustness for complex, tangled structures.
Specifically, we maintain $\mathcal{O}=\{\alpha_e\}_{e=1}^{\vert\mathcal{E}\vert}$, where each $\alpha_e\in[0,1]$ represents the existence probability of the edge.

\begin{figure*}[t]
\centering
\includegraphics{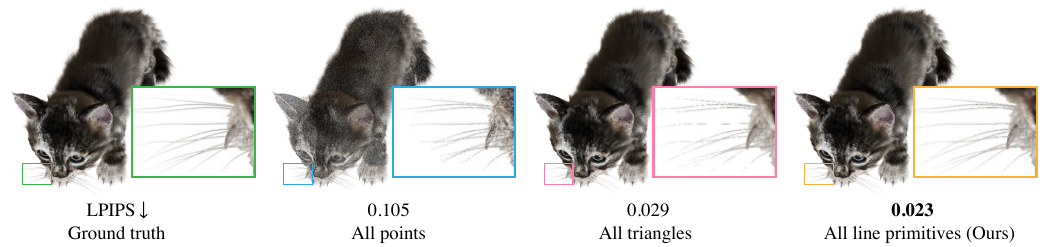}
\caption{
Comparison of different primitive types under the same 2M-vertex budget. (Left to right) Ground-truth view, point primitives, triangle primitives, and our line primitives. Point primitives are often too sparse, while triangles struggle with thin subpixel features; our line primitives utilize the vertex budget more effectively. Per-scene LPIPS scores are reported. See Section~\ref{sec:results} for further discussion.
}
\label{fig:ablations_prims}
\Description[Comparison of primitive types]{Comparison of primitive types}
\end{figure*}

\subsection{Differentiable Rasterization of Line Primitives}
We optimize these parameters by rendering the image $\mathbf{R}_t$ for the target view $t$ and computing the loss $\mathcal{L}(\mathbf{T}_t,\mathbf{R}_t)$ to obtain gradients.
For the $t$-th view, we first project the vertices into homogeneous camera space $\mathbf{v}^t_v\in\mathbb{R}^4$ and compute the vertex colors:
\begin{equation}
\label{eq:shading}
    \mathbf{C}_{v}^t = \mathcal{S}(\mathbf{v}_{v},\mathbf{A}_{v};\,t),
\end{equation}
where $\mathcal{S}$ is a user-defined, view-dependent shading function, such as spherical harmonics evaluation.
Next, each line primitive $\mathbf{e}_e=(v_1^e,v_2^e)$ generates fragments $f$ for all pixels covered by the line by interpolating the endpoint colors as
\begin{equation}
\label{eq:interp}
    \mathbf{C}_{f}^t = (1-w)\,\mathbf{C}^t_{v_1^e} + w\,\mathbf{C}^t_{v_2^e},
\end{equation}
where $w\in[0,1]$ is the perspective-correct interpolation weight at the pixel sampling point.
The depth $d_f^t$ of the fragment is interpolated analogously from the homogeneous vertex positions of the endpoints.
Finally, the fragment opacity is derived from the edge opacity as
\begin{equation}
\label{eq:prim_opacity}
    \alpha_f^t = \alpha_e.
\end{equation}
The resulting fragments are passed to the stochastic depth test~\eqref{eq:stoc_depth_test}, yielding a single stochastic rendering $\mathbf{R}_t$.

\paragraph*{Line width}
While interpolation along the segment is straightforward, determining line width is nontrivial.
Existing methods either use a world-space tube radius~\cite{Zhang2023Projective,Worchel2023parametric} or a screen-space stroke width~\cite{Li:2020:DVG,Tojo2024Wireart}, which can be optimized via rendering gradients.
However, these approaches typically target sparse, stroke-based representations, whereas we focus on dense arrangements of subpixel-width fibers that yield fine-scale appearance.
This introduces two key challenges.
First, both forward and backward rendering must be highly efficient to handle a large number of line primitives.
Second, optimizing line widths below one pixel is inherently underconstrained, often leading to instability or requiring excessive sampling to recover width from anti-aliased pixel colors, where undersampling produces artifacts such as dashed lines.

\begin{figure}[b]
\centering
\includegraphics{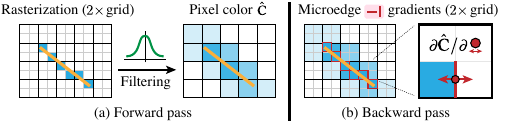}
\caption{%
Forward and backward passes of our Bresenham-based rasterization.
(a) We first rasterize binary line coverage on a $2\times$ supersampled grid, then apply Gaussian filtering to obtain anti-aliased pixel colors.
(b) During differentiation, gradients are propagated through microedges~\cite{Pidhorskyi2024rasterized} induced by line movements on the supersampled grid.
}
\label{fig:bresenham}
\Description[Our Bresenham-based rasterization]{Our Bresenham-based reasterization}
\end{figure}

To address these challenges, we adopt the classic Bresenham algorithm~\cite{bresenham}, which does not require an explicit line-width parameter.
Rather than explicitly parameterizing line width, we rasterize each line primitive on a $2\times$ subpixel grid and rely on subsequent anti-aliasing to produce a smooth, filtered appearance (Figure~\ref{fig:bresenham}(a)).
The Bresenham algorithm produces lines with consistent subpixel coverage, avoiding the instability associated with optimizing continuous line widths while still capturing fine-scale structure.
Moreover, its consistent coverage avoids undersampling artifacts such as dashed lines and is highly efficient, with widespread hardware support.
While we focus on this formulation for robustness and efficiency, our approach can be readily extended to alternative line rendering techniques, such as camera-facing quads commonly used in real-time graphics.

Bresenham rasterization produces blocky, axis-aligned fragments in screen space and lacks a smooth, continuous outline at the pixel level.
While these discretization effects are visually negligible after anti-aliasing and have little effect on rendering quality, they pose challenges for differentiation.
Specifically, they are incompatible with differentiable stroke rendering methods that rely on edge sampling~\cite{Li:2018:DMC,Li:2020:DVG} to compute positional gradients along the silhouettes of finite-width primitives.
To address this, we draw inspiration from the microedge gradient method~\cite{Pidhorskyi2024rasterized}, which treats rasterized fragment boundaries as proxies for triangle silhouettes.
We extend this idea to line primitives by treating the axis-aligned fragment edges produced by Bresenham rasterization as surrogate edges (Figure~\ref{fig:bresenham}(b)).
This formulation yields stable gradients, enabling effective optimization of width-free, inherently discrete line representations under anti-aliased rendering.

\paragraph*{Anti-aliasing}
After rasterization on a $2\times$ subpixel grid, we reconstruct the final image using multi-sample anti-aliasing (MSAA).
Different MSAA variants trade off quality and efficiency by varying the number and placement of samples.
To support real-time applications and efficient optimization of numerous primitives, we use fixed sample locations at the centers of a $2\times$ subpixel grid.
While fixed sampling can be sensitive to fine geometry, Bresenham-style single-pixel-width rasterization helps stabilize subpixel coverage, providing a good balance between efficiency and quality.

Various pixel reconstruction filters exist for aggregating subpixel samples into final pixel values.
Simple approaches include box filtering and filters based on distance to primitive edges~\cite{engel2014gpu}.
At runtime, these choices can be adjusted to balance visual quality and performance, or even omitted entirely for faster rendering.
During optimization, we use a high-quality filter to obtain smooth gradients and accurate geometry, while allowing flexible, lower-cost filtering at runtime.
Specifically, we use a Gaussian reconstruction filter with a standard deviation of $0.5$ pixels and a cut-off radius of $1.5$ pixels, implemented as a lightweight, differentiable image-filtering layer after rendering.

\begin{figure}[t]
\centering
\includegraphics{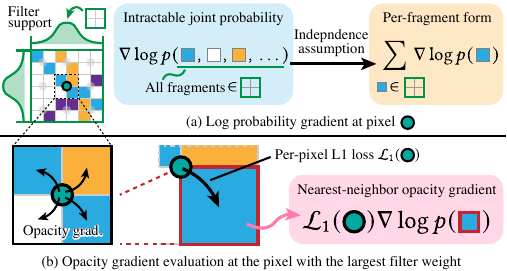}
\caption{%
Our opacity gradient formulation.
(a) We approximate the intractable joint probability within the filter support using a per-fragment formulation under an independence assumption.
(b) To stabilize optimization near fuzzy boundaries, per-fragment opacity gradients are evaluated only at the four fragments inside the pixel where the per-pixel L1 loss is computed.
}
\label{fig:opacity_grad}
\Description[Our opacity gradients]{Our opacity gradients}
\end{figure}

\paragraph*{Opacity gradients}
Unlike the traditional color gradient, the stochastic opacity gradient in~\eqref{eq:lr_gradient} requires dedicated treatment under anti-aliasing, where the pixel color is no longer determined by a single fragment, but depends on the joint contribution of multiple fragments within the filter kernel support $\mathcal{N}$.
Specifically, for an anti-aliased pixel color $\hat{\mathbf{C}}$, the opacity gradient in~\eqref{eq:lr_gradient} becomes
\begin{equation}
\label{eq:aa_opac}
    \nabla_{\alpha_f}\mathcal{L}
    =
    \mathbb{E}_{p_{\mathcal{N}}}
    \left[
        \mathcal{L}_1(\hat{\mathbf{C}})
        \nabla_{\alpha_f}\log p_{\mathcal{N}}
    \right],
\end{equation}
where $p_{\mathcal{N}}$ denotes the joint probability of all fragments $f\in\mathcal{N}$.
While the exact expression of $p_{\mathcal{N}}$ depends on the underlying primitive representation and is generally hard to obtain (Figure~\ref{fig:opacity_grad}(a)), assuming independent fragment probabilities yields
\begin{equation}
\label{eq:joint_indep}
    \nabla\log p_{\mathcal{N}}
    =
    \nabla\log
    \prod_{f\in\mathcal{N}} p_{f}
    =
    \sum_{f\in\mathcal{N}}
    \nabla\log p_{f}.
\end{equation}

Despite its practical per-fragment formulation, \eqref{eq:joint_indep} requires care when used for opacity updates.
First, fragments are not fully independent when they originate from a common primitive, which has a constant opacity inside.
The per-fragment gradients can produce inconsistent opacity gradients within a primitive, especially near the object silhouette.
For example, the gradient may encourage low opacity around one end of a line primitive while keeping it high around the other end, effectively shortening the line.
Fortunately, such \emph{geometric} updates are handled by positional gradients (i.e., by moving vertices).
Hence, we average the gradient over a primitive, allowing the opacity gradient to focus on the overall existence of a primitive, instead of its shape.

Second, since \eqref{eq:joint_indep} does not explicitly consider the anti-aliasing filter, each stochastic rasterization produces highly noisy opacity gradients for fragments with small filter weights.
Such fragments with small filter weights contribute little to the pixel color and loss, making it difficult to determine appropriate opacity updates from a single stochastic rasterization.
Hence, we evaluate the opacity gradient only for the fragments with the largest filter weights (i.e., four immediate neighbor fragments of the pixel centers) as
\begin{equation}
\label{eq:subgrid_opac}
    \nabla\log p_{\mathcal{N}}
    \approx
    \sum_{f\in\mathcal{N}_{\mathrm{nearest}}}
    \nabla\log p_{f},
\end{equation}
where $\mathcal{N}_{\mathrm{nearest}}$ denotes the $2\times 2$ fragments inside the pixel where the color is evaluated at its center (see Figure~\ref{fig:opacity_grad}(b)).
This means that our opacity formulation assumes a $6\times 6$ Gaussian filter in the forward path and a $2\times 2$ box filter in the backward path.
Together with the full positional gradients and our below-mentioned re-lining operations, we find that this strategy reliably converges to fine geometry even from an unstructured initialization of primitive locations and orientations.
In practice, we implement the opacity gradient by nearest-neighbor upsampling of the loss image and applying the gradient formula~\eqref{eq:lr_gradient} on the $2\times$ grid.

\begin{figure}[b]
\centering
\includegraphics{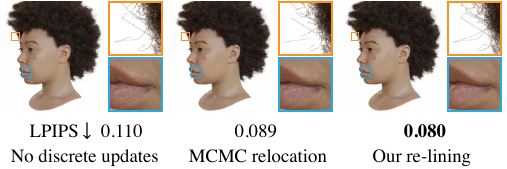}
\caption{%
Effect of re-lining. (Left to right) Reconstruction without discrete updates, with Markov Chain Monte Carlo (MCMC) primitive relocation~\cite{kheradmand20243d}, and with our re-lining strategy. Per-scene LPIPS scores against the ground-truth views are reported.
}
\label{fig:remesh}
\Description[Effect of re-lining]{Effect of re-lining}
\end{figure}

\subsection{Optimization and Implementation}

We adopt the photometric loss of~\cite{kerbl3Dgaussians} defined as
\begin{equation}
\mathcal{L}_{\text{photo}} = \lambda\,\mathcal{L}_{\text{D-SSIM}} + (1-\lambda)\,\mathcal{L}_1,
\end{equation}
with $\lambda=0.8$.
This loss is used to compute positional and color gradients, while the per-pixel $\mathcal{L}_1$ term with unit weight is used to evaluate the stochastic log-probability gradient in~\eqref{eq:log_prob_grad}.
We also apply a weak regularization $\mathcal{L}_{\text{len}}$ based on the average squared edge length to facilitate the connectivity adaptation introduced next.

\paragraph*{Adaptive re-lining}
To adapt connectivity during optimization, we periodically perform discrete updates, termed \emph{re-lining}.
To remove redundancy, we prune edges with low opacity or short length using thresholds.
We also merge polyline endpoints within the length threshold, forming connected polylines.
After pruning and merging, long edges are split to reach the target vertex count, as vertex attributes often dominate memory footprint.
This strategy utilizes the vertex budget more efficiently than approaches without shared segment endpoints, such as~\cite{kheradmand20243d} (Figure~\ref{fig:remesh}).

\paragraph*{Optimization setup}
Although our method can operate from fully random initialization, we first extract a coarse proxy surface using NeuS2~\cite{neus2} to accelerate convergence.
We then densely initialize disjoint edges around the surface to match the target vertex count (typically 1--2M).
It is more effective to let primitives \emph{emerge} through optimization than to rely on extensive positional updates or pruning.
Accordingly, we use very short initial edge lengths based on nearest-neighbor distances and initialize opacity to a small value of 0.1.
We use aggressive re-lining thresholds set to half their respective initial values to encourage topology adaptation.
We optimize vertex positions using VectorAdam~\cite{ling2022vectoradam} and all other parameters using Adam~\cite{kingma2014adam}.
Optimization is performed for 50{,}000 iterations, rendering one image per iteration.
Re-lining is applied every 100 iterations until iteration 35{,}000. 
The initial pruning threshold is used until iteration 25{,}000, after which the opacity threshold is linearly ramped to the test-time value of 0.5.
Full initialization details, hyperparameters, and schedules are provided in the supplementary material.

\paragraph*{Implementation}
Our differentiable rendering method is implemented using Vulkan~\cite{vulkan_spec} for rasterization and exposed as a PyTorch~\cite{paszke2019pytorch} operator via nanobind~\cite{nanobind}.
Forward rendering is performed entirely within vertex and fragment shaders, where random variables $\tau_f$ are generated in the fragment shader using a counter-based random number generator~\cite{salmon2011parallel}.
All fragments of the same primitive share the same counter value to avoid spurious holes within the primitive.
The opacity gradient is computed in a separate rendering pass using a custom fragment shader.
Color and positional gradients based on~\cite{Pidhorskyi2024rasterized} are computed in a compute shader.

%% file: sec/5_results.tex
\begin{figure}[b]
\centering
\includegraphics{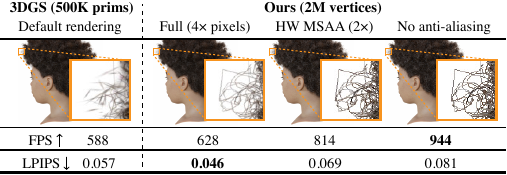}
\caption{%
Rendering speed comparison on the \textit{Shelly} dataset. FPS and LPIPS are averaged over 6 scenes. Even with full-quality rendering using $4\times$ more pixels, our method remains faster than 3DGS while achieving better perceptual quality. Since our representation reconstructs explicit line geometry, rendering quality and performance can be flexibly traded off using different anti-aliasing modes, including efficient hardware MSAA (HW MSAA) or disabling anti-aliasing entirely for maximum rendering speed.
}
\label{fig:shelly_runtime}
\Description[Rendering speed]{Rendering speed}
\end{figure}

\section{Results}
\label{sec:results}

We evaluate our method by reconstructing line-based representations from multi-view images.
We use a synthetic dataset to compare view-synthesis fidelity against relevant baselines, and real-world captures to demonstrate that our method successfully reconstructs highly challenging geometries and visual appearance encountered in practice.
All experiments are conducted on a workstation equipped with an Intel Core i9-14900K CPU, 64~GB of RAM, and an NVIDIA GeForce RTX 4090 GPU with 24~GB of VRAM.
Unless otherwise specified, our method uses 2M vertices with per-vertex Spherical Harmonics (SH) up to degree 3, and optimization takes approximately 45 minutes per example.

\begin{table}[t]
\centering
\small
\setlength{\tabcolsep}{5pt}
\caption{%
Rendering quality metrics on the \textit{Shelly} dataset, averaged over 6 scenes. $^\dagger$Results reported from the original paper~\cite{adaptiveshells2023}.
We also report the total memory footprint (Mem.) and the number of geometry parameters per primitive (Geom.), i.e., positions, topology, and Gaussian transformation parameters. Non-integer values reflect vertex sharing.
}
\label{tab:shelly_visual}
\begin{tabular}{lccccc}
\toprule
Method
& PSNR $\uparrow$
& SSIM $\uparrow$
& LPIPS $\downarrow$
& Mem. $\downarrow$
& Geom. $\downarrow$ \\
\midrule
DiffSoup        & 30.63 & 0.920 & 0.112 & \textbf{68\,MB} & 9 \\
VolSurfs        & 34.68 & 0.933 & 0.109 & 116\,MB & 6.02 \\
AdaptiveShells$^\dagger$ & \third{36.02} & \third{0.954} & \third{0.079} & - & - \\
3DGS           & \first{37.73} & \second{0.960} & \second{0.057} & 118\,MB & 10 \\
\midrule
Ours & \second{37.09} & \first{0.963} & \first{0.046} & 409\,MB & \textbf{3.59} \\
\bottomrule
\end{tabular}
\end{table}

\begin{figure*}[t]
\centering
\includegraphics{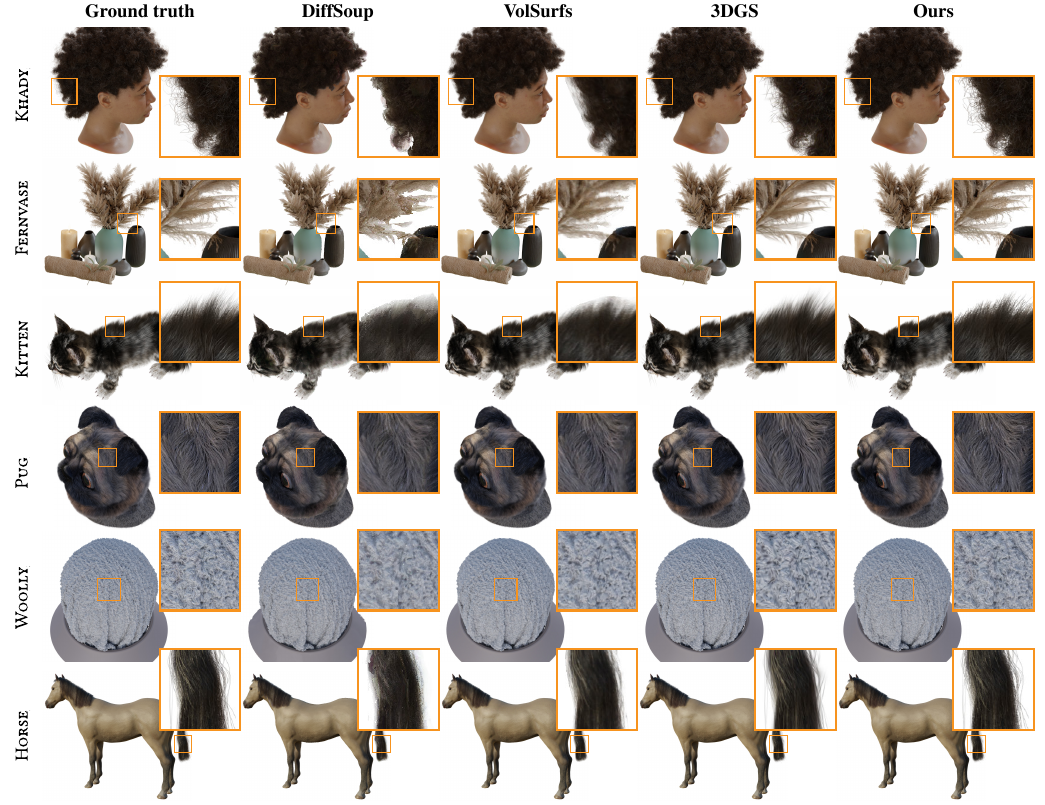}
\caption{%
Qualitative results on the \textit{Shelly} dataset. Rows correspond to different scenes, and columns show the ground truth and reconstruction results for each method. Compared to the surface-based methods DiffSoup and VolSurfs, our method better preserves thin fuzzy boundaries and fine geometric details. Compared to 3DGS, our method more faithfully captures connected curve structures while achieving comparable overall visual quality.
}
\label{fig:shelly}
\Description[Qualitative results on the Shelly dataset]{Qualitative results on the Shelly dataset}
\end{figure*}

\paragraph*{Synthetic dataset}
We first evaluate our method on the \textit{Shelly} dataset~\cite{adaptiveshells2023}, which provides path-traced renderings of furry geometries for novel-view synthesis evaluation.
The 3D models in the dataset were created by professional artists using a dedicated mixture of geometric primitives, including curves and textured surfaces.
We compare against both surface-based and volumetric approaches.
Among surface-based methods, we compare against DiffSoup~\cite{tojo2026diffsoup}, which optimizes a compact set (15K) of textured triangles without explicit anti-aliasing, and VolSurfs~\cite{Esposito2025VolSurfs}, which represents fuzzy geometry using layers of semi-transparent surfaces.
For volumetric methods, we compare against AdaptiveShells~\cite{adaptiveshells2023} using the quantitative results reported in their paper, as the official implementation is not publicly available.
Lastly, we compare against 3DGS~\cite{kerbl3Dgaussians}, where we control the number of Gaussian primitives using the Markov Chain Monte Carlo (MCMC) strategy~\cite{kheradmand20243d}.
We limit the number of Gaussian primitives in the 3DGS baseline to 500K, which yields a rendering speed comparable to that of our method (Figure~\ref{fig:shelly_runtime}).

\begin{figure*}[t]
\centering
\includegraphics{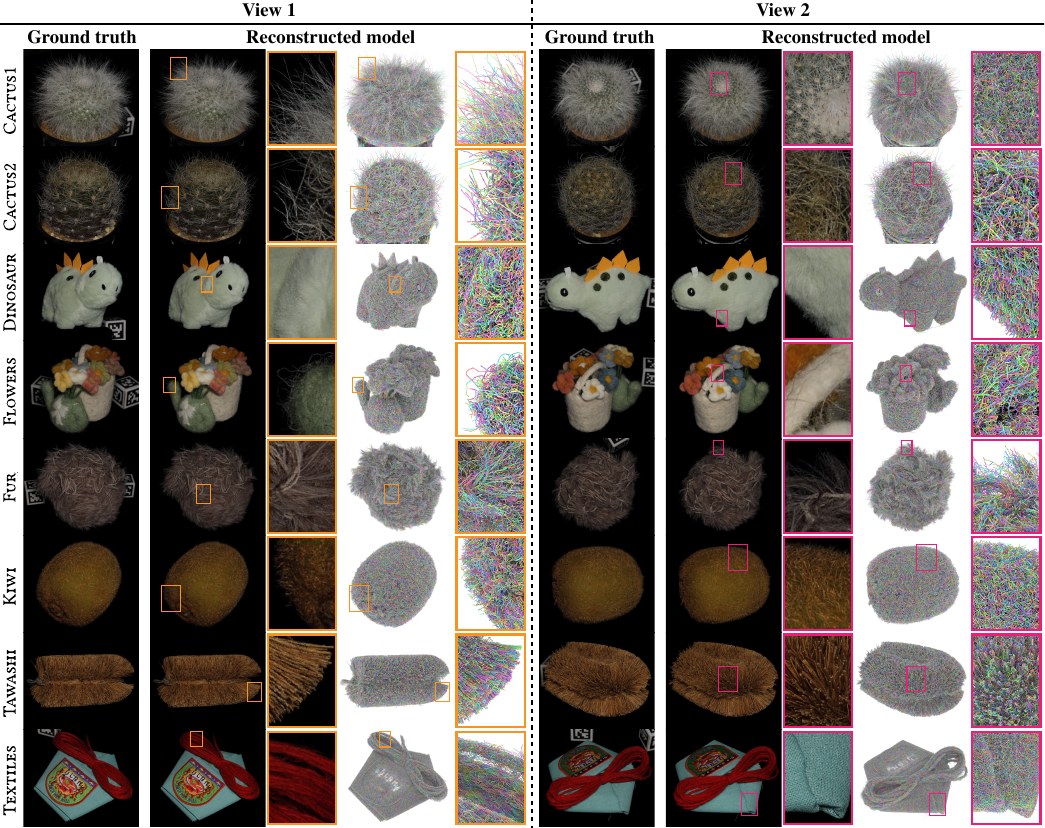}
\caption{%
Reconstruction results on our real-world capture dataset. Each row corresponds to a different captured object, while the left and right halves show two representative captured views used for optimization. For each view, we visualize the ground-truth image, the rendering, and the explicit line geometry of the reconstructed model. Uncompressed original images of the references, renderings, and geometry visualizations are provided in the supplemental material.
}
\label{fig:capture_results}
\Description[Reconstruction results on the real-world dataset]{Reconstrcution results on the real-world dataset}
\end{figure*}

Table~\ref{tab:shelly_visual} summarizes the quantitative results, and Figure~\ref{fig:shelly} shows qualitative reconstructions.
Our method achieves significantly better results than surface-based approaches, particularly in capturing complex anisotropic boundary shapes.
Although DiffSoup also relies on stochastic rasterization of opaque primitives, it struggles to reproduce fuzzy boundaries because it does not explicitly account for anti-aliasing.
Interestingly, our line-based representation performs well even on surface regions (e.g., the body of the \textsc{Horse} model), as it captures fine textures through the dense aggregation of subpixel line primitives.
Furthermore, our method achieves performance comparable to volumetric methods, with similar or slightly lower PSNR but substantially better scores on perceptually aligned metrics such as SSIM and LPIPS.
This suggests that our representation is more effective at capturing sharp, explicit geometry, albeit with a slight reduction in reproducing the extremely smooth boundary appearance in the ground-truth renderings generated with a very high number of samples per pixel.
Finally, although we do not optimize the total memory footprint, our polyline-based representation is parameter-efficient (Table~\ref{tab:shelly_visual}, Geom.).
By sorting vertices along the polylines, the 1D topology is encoded solely by the polyline break positions.
Depending on the application, memory can be reduced by using lower-order SH or other vertex attributes.

Note that the original \textit{Shelly} assets were authored using curves but require highly specialized modeling workflows, procedural techniques, and probably hours to days of substantial manual effort, whereas our method reconstructs similarly complex structures directly from multi-view images through optimization alone within one hour.
This advantage becomes even more apparent in the challenging real-world captures.

\paragraph*{Real-world dataset}
We capture a range of real-world objects exhibiting fuzzy, fiber-like geometry, including plants, fabrics, fur, and fruit, using the controlled setup summarized in Figure~\ref{fig:workflow}(a).
We name this dataset \textit{Fuzzy} and make it publicly available.
After image capture in a dark environment, we extract a coarse foreground mask by thresholding low-intensity pixels, applying morphological closing to preserve thin structures, and selecting the image-space connected component with the highest total saturation.
This preprocessing step mitigates the influence of low-intensity gray background pixels on our inverse rendering workflow, which assumes a fully black background.

Figure~\ref{fig:capture_results} shows line-based models obtained using our method, including rendered results and geometry visualizations from two captured views used for optimization.
As demonstrated, differentiable rendering of opaque line primitives with explicit anti-aliasing can recover a wide range of real-world geometries with varying fuzziness, fiber density, length, and spatial complexity.
The resulting models not only faithfully reproduce detailed textures and fuzzy boundaries, but also recover highly plausible outlines and directional structures of fiber-like elements that would be extremely difficult to author manually.
The reconstructed models can be rendered interactively with smoothly varying viewpoints on a laptop (M5 MacBook Air with 32~GB unified memory), as shown in the supplemental video.
While our method does not guarantee fully connected polyline structures that exactly follow the underlying physical connectivity, the reconstructed geometry nevertheless captures coherent explicit structures despite the limitations of casual captures, including occlusions and defocus blur.
These results open new possibilities for geometric workflows that require explicit structural representations beyond unstructured volumetric representations.

\begin{figure}[t]
\centering
\includegraphics{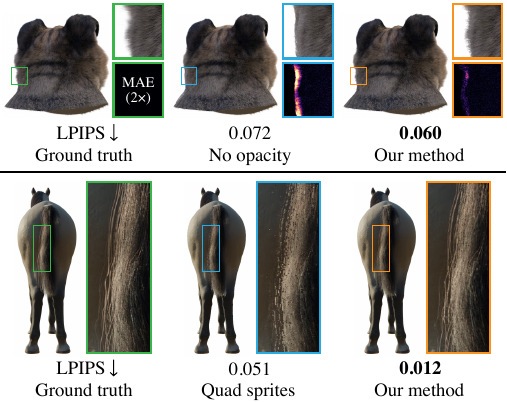}
\caption{
Ablation on our differentiable rendering components.
(Top, left to right) Ground-truth view, result without opacity optimization, and result using our method.
Close-ups show per-pixel mean absolute error (MAE), visualized with the magma colormap at $2\times$ magnification.
(Bottom, left to right) Ground-truth view, result using quad-based line rasterization with a 1~mm world-space line width, and result using our Bresenham-based rasterization with MSAA.
Per-scene LPIPS scores are reported.
}
\label{fig:ablations_diffrend}
\Description[Ablation on differentiable rendering components]{Ablation on differentiable rendering components}
\end{figure}

\begin{figure}[b]
\centering
\includegraphics{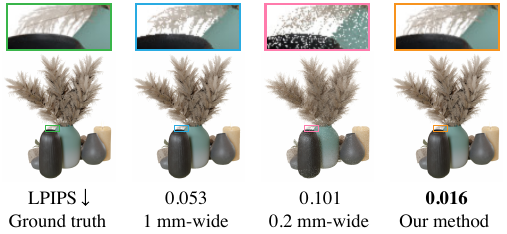}
\caption{%
Additional comparison with line rasterization using camera-facing quad sprites. (Left to right) Ground-truth view, result using quad-based rasterization with 1~mm world-space line width, result using 0.2~mm world-space line width, and our result using Bresenham-based rasterization with MSAA. Per-scene LPIPS scores are reported.
}
\label{fig:quads}
\Description[Comparison of quad-based and Bresenham-based line rasterization]{Comparison of quad-based and Bresenham-based line rasterization}
\end{figure}

\paragraph*{Applications}
Figure~\ref{fig:applications} shows various applications of our line-based representations.
The recovered explicit line primitives support applications such as hair reflectance rendering~\cite{marschner2003} and mass-spring animation, where additional KNN edges ($K=6$) connect the reconstructed polylines.
The simulation is solved on the GPU using Chebyshev-accelerated Jacobi projective-dynamics steps~\cite{Bouaziz14projective,wang16elastic}, processing 2M vertices in $\sim$55~ms/frame.
While the explicit representation is useful for physical simulation, handling millions of fibers in real time remains challenging and requires future investigation.
Our line-based rasterization remains efficient enough to support interactive preview on a mobile device.
Furthermore, our explicit line primitives can be exported as constant-radius world-space cylinders for offline ray-traced rendering.
The exported cylinders can be integrated with surface-based scenes under complex light transport using GPU-accelerated ray-primitive intersection queries provided by Mitsuba~3~\cite{Mitsuba3}.

\paragraph*{Ablation studies}
We also conduct ablation studies on our key design choices.
Figure~\ref{fig:ablations_diffrend} compares our differentiable rendering components with those used in traditional differentiable line-rasterization approaches (e.g.,~\cite{Takimoto_2024_CVPR}).
With an unstructured initialization, removing opacity optimization results in inaccurate boundary structures, highlighted by the per-pixel error map.
Rasterizing thin camera-facing quad sprites on the original pixel grid introduces severe undersampling artifacts, which are effectively resolved by our Bresenham-based rasterization with MSAA.
Figure~\ref{fig:quads} further analyzes the influence of line width in quad-based line rasterization.
In addition to a 1~mm world-space line width, we show results using 0.2~mm, as used for hair reconstruction in~\cite{Takimoto_2024_CVPR}.
While suitable for controlled hair capture, such thin lines exacerbate undersampling on general fuzzy objects, leading to overly sparse reconstructions with many spurious holes.
In contrast, our Bresenham-based line rasterization requires no world-space line width and consistently produces higher-quality results.
Note that scenes in the \textit{Shelly} dataset~\cite{adaptiveshells2023} are normalized independently; these line widths are therefore relative to the normalized scene extent rather than physical dimensions (approximate per-scene extents are reported in the supplemental material).
Nevertheless, this comparison illustrates the difficulty of selecting an appropriate line width without strong assumptions about object scale or the capture setup, highlighting the robustness of our method across diverse fuzzy objects.

\begin{table}[t]
\centering
% \small
\setlength{\tabcolsep}{5pt}
\caption{%
Quantitative results for the re-lining strategy ablation on the \textit{Shelly} dataset, averaged over 6 scenes.
}
\label{tab:remesh}
\begin{tabular}{lccc}
\toprule
Method              & PSNR $\uparrow$ & SSIM $\uparrow$ & LPIPS $\downarrow$ \\
\midrule
No discrete updates & \third{35.48}   & \third{0.949}   & \third{0.065}      \\
MCMC relocation     & \second{36.84}  & \second{0.960}  & \second{0.051}     \\
\midrule
Ours       & \first{37.09}   & \first{0.963}   & \first{0.046}      \\
\bottomrule
\end{tabular}
\end{table}

\begin{table}[t]
\centering
% \small
\setlength{\tabcolsep}{5pt}
\caption{%
Quantitative comparison with traditional line rasterization approaches on the \textit{Shelly} dataset, averaged over 6 scenes.
}
\label{tab:diffrend}
\begin{tabular}{lccc}
\toprule
Method                  & PSNR $\uparrow$ & SSIM $\uparrow$ & LPIPS $\downarrow$ \\
\midrule
No opacity        & \second{35.12}  & \second{0.958}  & \second{0.053}     \\
1~mm-wide quads   & \third{31.62}   & \third{0.919}   & \third{0.111}      \\
0.2~mm-wide quads & 22.52           & 0.801           & 0.178              \\
\midrule
Ours     & \first{37.09}   & \first{0.963}   & \first{0.046}      \\
\bottomrule
\end{tabular}
\end{table}

\begin{figure}[b]
\centering
\includegraphics{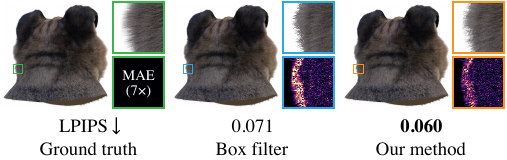}
\caption{%
Ablation of the Gaussian reconstruction filter used for MSAA. (Left to right) Ground-truth view, result using a box filter on the $2\times$ subpixel grid, and our result using a Gaussian filter on the same grid. Close-ups show per-pixel mean absolute error (MAE), visualized with the magma colormap at $7\times$ magnification. Per-scene LPIPS scores are reported.
}
\label{fig:filters}
\Description[Comparison of box and Gaussian reconstruction filters]{Comparison of box and Gaussian reconstruction filters}
\end{figure}

\begin{table*}[t]
\centering
% \small
\setlength{\tabcolsep}{5pt}
\caption{%
Per-scene LPIPS scores on the \textit{Shelly} dataset for all ablated configurations (lower is better). ``Re-lining'' corresponds to Figure~\ref{fig:remesh}, ``Differentiable rendering components'' to Figures~\ref{fig:ablations_diffrend},~\ref{fig:quads}, and~\ref{fig:filters}, and ``Primitive types'' to Figure~\ref{fig:ablations_prims}. Ties in the ranking are resolved using unrounded values.
}
\label{tab:ablation_full}
\begin{tabular}{lccccccccc}
\toprule
& \multicolumn{2}{c}{Re-lining} & \multicolumn{4}{c}{Differentiable rendering components} & \multicolumn{2}{c}{Primitive types} &  \\
\cmidrule(lr){2-3}\cmidrule(lr){4-7}\cmidrule(lr){8-9}
Scene             & No discere  & MCMC    & No opacity     & 1~mm quads & 0.2~mm quads & Box filter        & Points & Triangles     & Ours  \\
\midrule
\textsc{Khady}    & 0.110 & \second{0.089} & \third{0.089}  & 0.131       & 0.221         & 0.099         & 0.247  & 0.106         & \first{0.080}  \\
\textsc{Fernvase} & 0.026 & \second{0.017} & 0.020          & 0.053       & 0.101         & 0.021         & 0.081  & \third{0.018} & \first{0.016}  \\
\textsc{Kitten}   & 0.039 & \second{0.026} & \third{0.027}  & 0.071       & 0.116         & 0.035         & 0.105  & 0.029         & \first{0.023}  \\
\textsc{Pug}      & 0.080 & \second{0.064} & 0.072          & 0.147       & 0.228         & \third{0.071} & 0.210  & 0.071         & \first{0.060}  \\
\textsc{Woolly}   & 0.112 & \third{0.095}  & 0.100          & 0.213       & 0.293         & 0.109         & 0.256  & \first{0.066} & \second{0.084} \\
\textsc{Horse}    & 0.020 & \third{0.013}  & \second{0.012} & 0.051       & 0.108         & 0.015         & 0.099  & 0.014         & \first{0.012}  \\
\midrule
Average           & 0.065 & \second{0.051} & 0.053          & 0.111       & 0.178         & 0.059         & 0.166  & \third{0.051} & \first{0.046}  \\
\bottomrule
\end{tabular}
\end{table*}

Figure~\ref{fig:ablations_prims} compares different primitive types optimized by our differentiable rasterizer.
Point primitives are too sparse to reproduce dense fiber structures, while triangle primitives provide smooth interpolation but struggle to capture thin subpixel features, leading to artifacts such as dashed whiskers.
Our explicit line primitives preserve coherent directional structures, fine-scale details, and smooth appearance transitions.
Figure~\ref{fig:filters} provides an ablation on the Gaussian reconstruction filter used for anti-aliasing.
Using a straightforward box filter on the same $2\times$ subpixel grid fails to reproduce smooth, fuzzy boundaries, resulting in higher rendering error and less coherent directional structures.
Our Gaussian reconstruction filter yields smoother, fuzzier boundaries while better preserving fine directional structures.

We provide additional quantitative results for the key ablation studies motivating our inverse rendering workflow.
Table~\ref{tab:remesh} reports average rendering quality metrics over the \textit{Shelly} dataset for the re-lining ablation in Figure~\ref{fig:remesh}.
Table~\ref{tab:diffrend} reports corresponding metrics for the comparison of differentiable rendering components in Figures~\ref{fig:ablations_diffrend} and~\ref{fig:quads}.
Our method consistently outperforms all baseline configurations across all metrics.
In addition, Table~\ref{tab:ablation_full} reports per-scene quality scores for all ablations.
We report the perceptually aligned LPIPS metric.
Our method outperforms all baseline configurations across all scenes, except for the \textsc{Woolly} scene reconstructed using only triangle primitives.
Overall, the results support our use of line primitives and Bresenham-based rasterization, as further suggested by the competitive performance of the ``MCMC'' variant, which differs only in the discrete update strategy while sharing the same differentiable renderer as our full configuration.

\begin{figure}[b]
\centering
\includegraphics{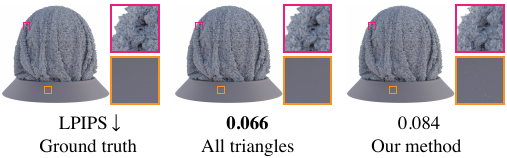}
\caption{%
Further analysis of the challenging \textsc{Woolly} scene, where the triangle-based representation outperforms our line-based method. (Left to right) Ground-truth view, result using only triangle primitives, and our result using only line primitives. Both representations use the same 2M-vertex budget. Per-scene LPIPS scores are reported.
}
\label{fig:limitation}
\Description[Triangle vs. line primitive comparison]{Triangle vs. line primitive comparison}
\end{figure}

\paragraph*{Failure cases}
Figure~\ref{fig:limitation} further analyzes the \textsc{Woolly} scene, the only scene where a triangle-based representation outperforms our line-based method.
Compared to other scenes in the \textit{Shelly} dataset, this scene spans a larger spatial extent and exhibits an overall rounded shell shape, making it difficult to sufficiently cover with line primitives.
Triangle primitives fill the object surface more efficiently, producing sharper contrast in textured regions while smoothly representing visually flat regions.
In contrast, our representation produces slightly fuzzier wool boundaries than the ground truth and occasionally reveals background pixels in visually flat regions, while preserving finer-scale fiber texture.
This example motivates a \emph{hybrid} representation as a promising direction for future work: using triangle primitives for large, visually flat regions and deeper surface layers, and line primitives for texture-rich regions and fuzzy front boundaries, thereby combining the strengths of both representations.
Other failure cases arise from imperfect background removal, which can produce inaccurate reconstructed boundaries.
Our method can also interpret black surface regions as holes, as observed in the text region of the \textsc{Textile} model.

%% file: sec/6_conclusion.tex
\section{Limitations and Future Work}
\label{sec:conclusion}

This work has several limitations and directions for future work.
Recovering complete topology, including heavily occluded structures such as hair interiors, likely requires stronger priors or richer capture setups.
Our method does not recover fully accurate microscale fiber orientations and connectivity for structures that are barely visible at the pixel scale. Future work could replace hand-crafted connectivity update rules with learned directional connection probabilities that guide opacity optimization.
A more principled variance-reduction approach for opacity gradients, for example using the control variates method that incorporates Gaussian filter weights and primitive orientations, remains a natural extension.
Recovering line width from subpixel observations is fundamentally challenging but would also be valuable.
Finally, adapting the representation across viewing distances is a promising direction.
For example, avoiding visible line primitives in close-up views may require level-of-detail techniques or a hybrid triangle--line representation.
Overall, we believe our approach opens new possibilities for reconstructing and representing complex fuzzy appearance using explicit geometry, enabling broader integration with geometry-centric workflows.

%% file: sec/X_suppl.tex
% =============================================================
% Supplementary Material — formatting preamble
% =============================================================
\clearpage
\setcounter{section}{0}
\setcounter{figure}{0}
\setcounter{table}{0}
\setcounter{equation}{0}
\setcounter{algocf}{0}
\renewcommand{\thesection}{\Alph{section}}
\renewcommand{\thefigure}{A\arabic{figure}}
\renewcommand{\thetable}{A\arabic{table}}
\renewcommand{\theequation}{A\arabic{equation}}
\renewcommand{\thealgocf}{A\arabic{algocf}}

\section*{Supplemental Material: Inverse Rendering for Modeling with Line Primitives}

This supplemental material provides additional details on our real-world capture dataset and optimization settings.

\section{Additional Capture Details}

Table~\ref{tab:fuzzy_views} summarizes the statistics of our \textit{Fuzzy} dataset.
Given camera poses estimated by COLMAP~\cite{schonberger2016structure}, we reconstruct a coarse surface using NeuS2~\cite{neus2} to initialize line primitives around it.
Although NeuS2 efficiently recovers a coarse object outline within approximately three minutes, it often produces noisy signed distance fields in low-texture black background regions, leading to cluttered surfaces extracted with the marching cubes algorithm~\cite{Lorensen87mc}.
To mitigate this, we apply a simple manual preprocessing step that roughly aligns a bounding cube to the object and crops the reconstructed mesh by mesh intersection.
This is performed once per scene and prevents noisy background regions from affecting the initial line distribution, particularly the line lengths initialized from nearest-neighbor distances between the seed points.
We will additionally release these coarse proxy surfaces.
Note that cube intersection may not fully remove noisy floating surface components.
We allow the residual noise, as it can later be removed via our opacity optimization.

\section{Additional Optimization Details}

\paragraph*{Initialization}
After reconstructing the coarse surface, we uniformly sample seed points within a distance band of $0.1$.
We determine the band distance of $0.1$ experimentally and use it for all datasets.
Rather than concentrating samples tightly around the reconstructed surface, we intentionally use a conservatively wide band to avoid missing fine structures that surface reconstruction methods often fail to recover when applied to fuzzy geometry.
To efficiently perform rejection sampling within this distance band, we first evaluate distances on a coarse voxel grid and then apply stratified sampling to generate candidates with a higher acceptance rate than ambient-space sampling.
This strategy enables the sampling of 1M--2M points within a few seconds.
We note that scenes in the \textit{Shelly} dataset are normalized independently; although their normalized extents are similar, distances in the normalized coordinate system may not correspond directly to physical dimensions. For reference, Table~\ref{tab:mesh_bbox} reports the bounding-box extents of the coarse surfaces for all \textit{Shelly} scenes.

\paragraph*{Hyperparameters and scheduling}
Unless specified otherwise, we use learning rates similar to those commonly used in 3D Gaussian splatting~\cite{kerbl3Dgaussians}.
Specifically, we set the zeroth-order SH, higher-order SH residual, and line-opacity learning rates to $2.5{\times}10^{-3}$, $1.25{\times}10^{-4}$, and $5{\times}10^{-2}$, respectively.
For positional optimization, we found that stable convergence is more reliably achieved using consistently small learning rates rather than aggressive annealing schedules.
We therefore use initial positional learning rates that are 5--10$\times$ smaller than those commonly used in 3DGS: $2.5{\times}10^{-5}$ for the \textit{Shelly} dataset and $1.6{\times}10^{-5}$ for our \textit{Fuzzy} dataset, exponentially decayed to $1/5$ of their initial values during optimization.
The vertex-position learning rate is additionally scaled according to the camera extent, following standard 3DGS optimization practice.
The weight of the length regularization $\mathcal{L}_{\mathrm{len}}$ is set to $1.0\times10^{-3}$.

\paragraph*{Details of baseline methods}
For the 3DGS baseline~\cite{kerbl3Dgaussians}, we sample the initial seed points using the same distance-band sampling around the same coarse proxy surface as our method for a fair comparison.
While our method samples 1M seed points to initialize line primitives with 2M vertices, 3DGS directly uses 500K seed points to initialize 500K Gaussian primitives.
For the DiffSoup baseline~\cite{tojo2026diffsoup}, we initialize the triangle mesh using the MobileNeRF~\cite{chen2022mobilenerf} mesh, reported in the original paper to produce the best results on the \textit{Shelly} dataset.
VolSurfs~\cite{Esposito2025VolSurfs} relies on implicit surface reconstruction and therefore does not require external seed points or primitive initialization.

\begin{table}[t]
\centering
\caption{Per-scene statistics for our \textit{Fuzzy} dataset. All source images are captured at $6720\times4480$~px. COLMAP's image undistortion~\cite{schonberger2016structure} crops images slightly differently per scene, and optimization is performed on a $4\times$ downsampled version of the undistorted images.}
\label{tab:fuzzy_views}
\small
\begin{tabular}{lcccc}
\toprule
Scene & \#~Views & Orig.\ res. & COLMAP res. & Train res. \\
\midrule
\textsc{Cactus1}  & 173 & $6720\times4480$ & $6709\times4466$ & $1677\times1116$ \\
\textsc{Cactus2}  & 108 & $6720\times4480$ & $6712\times4471$ & $1678\times1117$ \\
\textsc{Dinosaur} & 129 & $6720\times4480$ & $6711\times4467$ & $1677\times1116$ \\
\textsc{Flowers}  & 122 & $6720\times4480$ & $6705\times4459$ & $1676\times1114$ \\
\textsc{Fur}      & 128 & $6720\times4480$ & $6720\times4474$ & $1680\times1118$ \\
\textsc{Kiwi}     & 124 & $6720\times4480$ & $6715\times4470$ & $1678\times1117$ \\
\textsc{Tawashi}  & 124 & $6720\times4480$ & $6712\times4470$ & $1678\times1117$ \\
\textsc{Textiles} & 156 & $6720\times4480$ & $6707\times4468$ & $1676\times1117$ \\
\bottomrule
\end{tabular}
\end{table}

\begin{table}[t]
\centering
\small
\setlength{\tabcolsep}{5pt}
\caption{%
Axis-aligned bounding-box extents of the coarse proxy meshes for the \textit{Shelly} dataset. We report the lengths of the shortest axis (Min.), longest axis (Max.), and the bounding-box diagonal (Diag.).
}
\label{tab:mesh_bbox}
\begin{tabular}{lcccccc}
\toprule
 & \textsc{Khady} & \textsc{Fernvase} & \textsc{Kitten} & \textsc{Pug} & \textsc{Woolly} & \textsc{Horse} \\
\midrule
Min.  & 1.14 & 1.24 & 0.92 & 2.23 & 2.99 & 0.80 \\
Max.  & 1.67 & 1.93 & 2.24 & 2.55 & 3.62 & 3.57 \\
Diag. & 2.45 & 2.84 & 2.81 & 4.20 & 5.93 & 4.49 \\
\bottomrule
\end{tabular}
\end{table}